\documentclass[aps,prl,reprint,amsmath,superscriptaddress,showpacs,amsfonts,amssymb,longbibliography]{revtex4-1}
\usepackage{graphicx}
\usepackage{bm}
\usepackage{color}
\usepackage[dvipsnames]{xcolor}
\usepackage{MnSymbol}
\usepackage{enumerate}
\usepackage{bbold}
\usepackage{lipsum}
\usepackage{array}
\usepackage{hyperref}
\usepackage{textgreek,wasysym}
\usepackage{natbib}
\usepackage{soul}

\usepackage{multirow}

\DeclareMathAlphabet\mathbfcal{OMS}{cmsy}{b}{n}

\def\be{\begin{equation}}
\def\ee{\end{equation}}

\newcommand{\beq}{\begin{equation}}
\newcommand{\eeq}{\end{equation}}
\newcommand{\nn}{\nonumber}

\newcommand{\erf}[1]{Eq.~(\ref{#1})}

\newcommand{\erfand}[2]{Eqs.~(\ref{#1}) and (\ref{#2})}

\newcommand{\smallfrac}[2]{\mbox{$\frac{#1}{#2}$}}

\newcommand{\half}{\smallfrac{1}{2}}
\newcommand{\bra}[1]{\langle{#1}|}
\newcommand{\ket}[1]{|{#1}\rangle}

\newcommand{\sq}[1]{\left[ {#1} \right]}

\newcommand{\ro}[1]{\left( {#1} \right)}

\newcommand{\tp}{^{\top}}

\renewcommand{\ul}[1]{\underline{{#1}}}
\renewcommand{\ss}{_\text{ss}}

\newcommand{\dd}{{\rm d}}

\newcommand{\kd}{\kappa}
\newcommand{\ks}{K}
\newcommand{\phis}{\Phi}
\newcommand{\phid}{\phi}
\newcommand{\cd}{c}
\newcommand{\xd}{X}

\newcommand{\ddt}{\tau} 
\newcommand{\noise}{\Xi}
\newcommand{\info}{Y}

\newcommand{\gu}{\gamma_\uparrow}
\newcommand{\gd}{\gamma_\downarrow}
\newcommand{\gud}{\gamma_{\uparrow,\downarrow}}
\newcommand{\bg}{\breve\gamma}
\newcommand{\mg}{\bar\gamma}
\newcommand{\rtp}{z}

\newcommand{\coh}{\mathcal{C}}
\newcommand{\cohc}{\coh^{\rm c}}
\newcommand{\cohnc}{\coh^{\rm nc}}
\newcommand{\ccc}{\mathcal{A}}

\newcommand{\opt}{^\star}

\definecolor{nblue}{rgb}{0.06,0.3,0.73}
\definecolor{nblack}{rgb}{0,0,0}
\definecolor{nred}{rgb}{0.9,0.1,0.1}
\definecolor{nmagenta}{rgb}{0.7,0.0,0.3}
\definecolor{neditcolor}{rgb}{0.3,0.3,0.9}

\newcommand{\blk}{\color{nblack}}

\newcommand{\letter}{Letter}
\newcommand{\sm}{CP}
\newcommand{\ie}{\emph{i.e.}}

\renewcommand{\section}[1]{\emph{#1}.\blk---}

\begin{document}

\title{Optimized mitigation of random-telegraph-noise dephasing \protect\\ by spectator-qubit sensing and control}
\author{Hongting Song}
\email{shtfc@163.com}
\affiliation{Centre for Quantum Computation and Communication Technology (Australian Research Council), \\ Centre for Quantum Dynamics, Griffith University, Yuggera Country, Brisbane, Queensland 4111, Australia}
\affiliation{Qian Xuesen Laboratory of Space Technology, China Academy of Space Technology, Beijing 100094, China}
\author{Areeya Chantasri}
\email{areeya.chn@mahidol.ac.th}
\affiliation{Optical and Quantum Physics Laboratory, Department of Physics, Faculty of Science, Mahidol University, Bangkok, 10400, Thailand}
\affiliation{Centre for Quantum Computation and Communication Technology (Australian Research Council), \\ Centre for Quantum Dynamics, Griffith University, Yuggera Country, Brisbane, Queensland 4111, Australia}
\author{Behnam Tonekaboni}
\email{behnam.tfn@gmail.com}
\affiliation{Centre for Quantum Dynamics, Griffith University, Yuggera Country, Brisbane, Queensland 4111, Australia}
\author{Howard M. Wiseman}
\email{h.wiseman@griffith.edu.au}
\affiliation{Centre for Quantum Computation and Communication Technology (Australian Research Council), \\ Centre for Quantum Dynamics, Griffith University, Yuggera Country, Brisbane, Queensland 4111, Australia}

\date{\today}

\begin{abstract}
Spectator qubits (SQs) are a tool to mitigate noise in hard-to-access data qubits. The SQ, designed to be much more sensitive to the noise, is measured frequently, and the accumulated results used rarely to correct the data qubits. For the hardware-relevant example of dephasing from random telegraph noise,  we introduce a Bayesian method employing complex linear maps which leads to a  plausibly optimal adaptive measurement and control protocol. The suppression of the decoherence rate is quadratic in the SQ sensitivity, establishing that the SQ paradigm works arbitrarily well in the right regime. 
\end{abstract}

\maketitle


Despite recent impressive advances towards large-scale quantum computing~\cite{AruAry2019,ZhoWan2020}, the challenge of suppressing noise sufficiently to achieve scalable universal quantum computing remains~\cite{TemBra2017,Pre2018,EndBen2018,KanTem2019}. The best known approaches to noise-mitigation are dynamical decoupling (DD), which works for non-Markovian noise~\cite{Vio99, viola2003robust, biercuk2011dynamical, NgLid2011, SouAlv2011, MedCyw2012, PazLid2013, ZhaSou2014}, and quantum encoding and error correction (QEC) which works best for Markovian and local noise~\cite{Shor1995,Steane1996,Terhal2015}.  In favourable regimes, both of these approaches  can suppress errors arbitrarily. 

For data qubits that are very well isolated from their environment, it could be difficult to control them (as for DD) or measure them (for QEC) rapidly. In this context, another paradigm for error mitigation has recently been proposed and demonstrated~\cite{GupEdm2020, MajAnd2020,SinBra2022}: spectator qubits (SQs). The SQ is located physically near to the data qubits. It is a spectator in two senses: it does not interact with the data qubits, but it is a sensitive probe to the noise they experience. The idea is that by measuring the SQ in a suitable way, the experimenter can obtain information about the noise in real time, and, by applying suitable controls, cancel at least some of the effect of that noise on the data qubits. 

Previous work within the SQ paradigm has used rather simple measurement and control strategies, and has not shown that the SQ can, like DD and QEC, work arbitrarily well in a suitable regime. In this \letter, we transform that situation. We consider an experimentally relevant type noise that effects many qubits: dephasing caused by a random telegraph process (RTP)~\cite{ItaTok2003, GalAlt2006, CulHu2009, BerGal2009}. We consider the regime where the flip-rates, $\gu$, $\gd$, of the RTP  are large compared to $\kd$, the data qubit's sensitivity to the RTP, but small compared to the SQ's sensitivity, $\ks$:
\begin{equation} \label{AsymReg}
T^{-1}, ~ \kd \ll \gud \ll \ks.
\end{equation}
Here $T$ is the time at which the control is applied on the data qubit. We introduce a principled method, based on Bayesian maps, to construct a measurement and control algorithm that is, we conjecture, optimal in this regime. Our algorithm suppresses the data qubit decoherence by an amount (\ie, a divisor) scaling as $(\ks/\mg)^2$, where $\mg := (\gu + \gd)/2$, limited only by the sensitivity of the SQ. 

We tackle the problem from the perspective of quantum estimation or  
decision theory~\cite{Helstrom,WisMil10}. The ultimate limits to such problems 
are surprisingly subtle even with a single probe qubit confined to a single plane~\cite{Acin2005,Higgins-discrim09,Higgins11,Slu17,vargas2021quantum,Ser11,Fer13,Shu14,Sek17}. In all of these, the optimal  sequence of single-qubit measurements is, in general, {\em adaptive}; that is, the basis for later measurements must depend on the results of earlier measurements~\cite{WisMil10,Ser11,Shu14}. The SQ problem we consider here is more complicated than the above examples~\cite{Acin2005, Higgins-discrim09, Higgins11, Slu17, vargas2021quantum, Ser11, Fer13, Shu14, Sek17} in three ways. First, it is dynamic: there is a Hamiltonian affecting the qubit rotation that changes stochastically (the RTP). Second, we allow a choice not just in the time of application of the Hamiltonian before measurement (as in~\cite{Ser11}), but also in the measurement basis. Third, the quantity to be estimated is not the current state of the SQ, nor even the current state of the RTP (\ie~the current Hamiltonian). Rather, it is the cumulative phase acquired by the data qubit, proportional to the integral of the RTP up to that time.

In this \letter, we show that, despite these complications, it is possible, in the relevant regime (\ref{AsymReg}), to derive a measurement and control strategy using the SQ that is plausibly optimal. That is, in the multiplicative factor by which the data qubit decoherence rate is reduced, 
\begin{align} \label{multfac}
H\opt \,(\mg/\ks)^2, \text{ where } H\opt \approx 1.254,
\end{align}
even the prefactor ($H\opt$) is as  small as possible. Unsurprisingly, we find that the optimized strategy is adaptive. 

Finding the ultimate limit to a harder qubit-probe estimation problem than has hitherto been considered serves as a benchmark against which other techniques, such as machine learning, can be compared. Having an optimal SQ protocol solution could help to hone heuristic algorithms for even more complicated problems. 

\begin{figure}
\includegraphics[width=8.5cm]{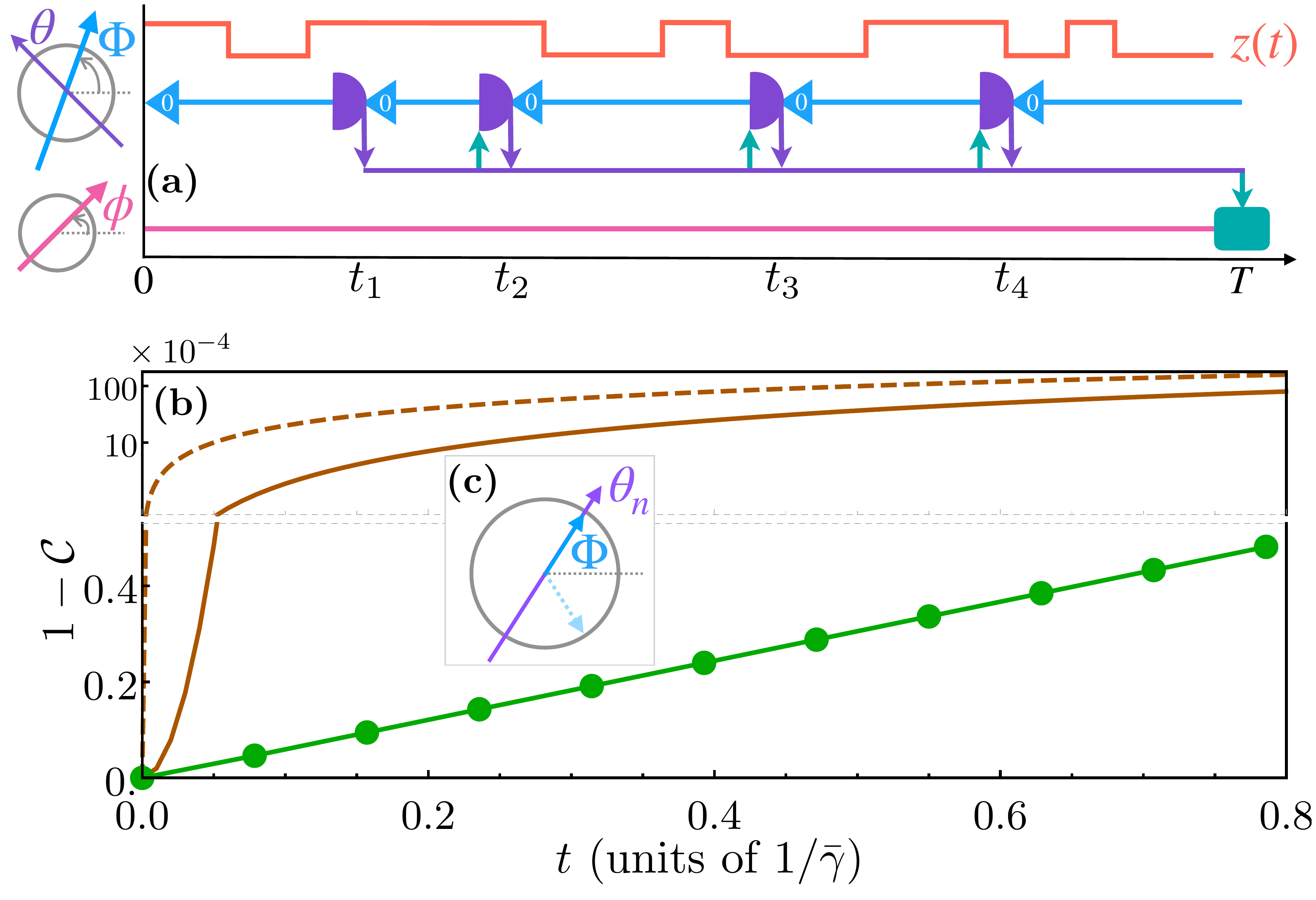}
\caption{(a)
Schema for the SQ protocol, showing dynamics of (from top to bottom): RTP (orange); SQ (blue) which is reprepared (triangle) after each measurement (purple semi-circle and downward arrows); measurement record (purple); and data qubit (pink). Teal arrows from measurement record show feedback (upwards for adaptive measurements on the SQ and downward to the teal box at $t = T$ representing the final control on the data qubit). Small Bloch spheres at left indicate the meaning of $\theta$, $\Phi$, and $\phi$. (b)~Plot of data qubit decoherence ($1-\coh$) versus $t$, with change to log-scale for $1-\coh > 0.55\times 10^{-4}$. From top to bottom: no-control asymptotic result, \erf{uncondec}, dashed brown; exact no-control case, solid  brown;  exact Greedy algorithm data points with fit (green). Parameters: $\kd=0.2, \gu=\gd = 1, \ks=20$. (c)~Bloch sphere showing how Greedy chooses the measurement axis $\theta_n$ to align with the most likely SQ state, $\ket{\Phi}^{\rm s}$.\label{fig1} 
}
\end{figure}

The structure of this paper is as follows. First we introduce the physical system: the RTP, the data qubit, and the SQ (see Fig.~\ref{fig1}). Next we introduce the calculational tool of Bayesian maps. Then we consider a greedy algorithm, and, with insights from that, construct the  optimized algorithm that achieves (\ref{multfac}), and verify it numerically. We conclude with open problems. More details in all sections are found in the companion paper (\sm)~\cite{PRA}. 

\section{Charge noise and RTP} 
One of the main causes of decoherence for solid-state qubits is the  
stochastic motion of particles in traps in oxide layers or interfaces ~\cite{ZorAhl1996,PalFao2002,ItaTok2003, GalAlt2006, CulHu2009, BerGal2009}.  The simplest model capturing the essential physics of such processes is the RTP $\rtp(t)$, which has a Lorentzian noise spectrum. The RTP switches between two values: $\rtp_t := \rtp(t)=\pm 1$. Defining 
$\underline{P}_t = \left(\wp(\rtp_t = +1), \wp(\rtp_t=-1)\right)\!\tp$
as the vector of probabilities at any given time, the master equation, and its steady state (ss) distribution, are, respectively~\cite{Gar85}
\beq  \label{MESS}
\ul{\dot{P}}_t = 
\left( \begin{matrix} -\gd  & +\gu  \\ +\gd & -\gu \end{matrix} \right) \ul{P}_t \;\; {\rm and} \;\; 
\ul{P}_{\rm ss} = \frac{1}{2 \mg} \left( \begin{matrix} \gu \\ \gd\end{matrix} \right) ,
\eeq
where $\mg = (\gu+\gd)/2$ as before. For the calculations below, we are also interested in the time-integrated noise,
\begin{equation} \label{IRTP}
\xd = \int_0^t \!\rtp(s){\rm d}s,
\end{equation}
where we omit the $t$-dependence when it is not needed. In particular, we need to solve not just for $\wp(z_t)$ via Eq.~\eqref{MESS}, but also for $\wp(X, z_t)$; see the \sm~\cite{PRA}.

\section{Data qubit dephasing} 
We assume the data qubit decoherence is caused by phase fluctuation. The coherence remaining at any time 
can thus be calculated by taking the absolute value of the average of the phasors:  
\beq
\coh = \left|\left\langle e^{i\phid(\noise)}\right\rangle_\noise\right|,
\label{eq:Coh}
\eeq
where $\phid$ is the data qubit's phase and $\Xi$ is all the variables on which $\phid$ depends. (We ignore the dependence on $\phid(0)$ by taking that to be $0$). In this work, we assume the RTP causes the phase fluctuation via the data (d) qubit's Hamiltonian $\hat H_{\rm d} = \frac{\kd}{2} \, \hat\sigma_z^{\rm d}\rtp(t) $. Thus $\Xi$ arises from the RTP plus any controls applied to the data qubit. 

In the absence of control, we simply have $\noise = {\xd}$ and $\phid({\xd}) = \kd\, {\xd}$, as the data qubit phase just accumulates. In the \sm~\cite{PRA}, we give a closed-form expression for the no-control (nc) coherence, and show that in the long-time limit  ($t\gg  \bg^{-1}$), the coherence decays exponentially: 
\beq \label{uncondec}
\tfrac{{\rm d}}{{\rm d}t}\cohnc  = - \Gamma^{\rm nc} \cohnc \;;\quad\Gamma^{\rm nc} = \kd^2\bg/2\mg^2.
\eeq
Here $\bg := 2\gu\gd/(\gu+\gd)$ is the harmonic mean of $\gu, \gd$.

Say the quantum data is required at some time. If we have additional information, $\info$, about $\xd$, at that time, then we can increase the coherence $\coh$, by {\em controlling} the data qubit conditioned on $\info$. Specifically, a unitary $\hat\sigma_z^{\rm d}$-rotation can add a phase correction, $\cd(\info)$, to $\phid$. This means that the data qubit phase depends on two variables, $\Xi = ({\xd},\info)$, via  $\phid({\xd},\info) = \kd{\xd} -\cd(\info)$. Of course these two variables are not independent, as $Y$ contains information about $X$. We show in the \sm~\cite{PRA} that  the optimal choice for the control is 
\begin{equation} \label{optctrl}
\cd(\info) = \arg  \ccc_{|\info},  \text{ where } \ccc_{|\info} := \left\langle e^{i\kd { \xd}}\right\rangle_{{ \xd}|\info}.
\end{equation}
That is, whatever information $Y$ exists, this choice of control maximizes the coherence (\ref{eq:Coh}). Therefore, if the qubit is needed at time $T$, the appropriate {\em reward function} for our problem is $\cohc(T)$, where~\cite{PRA}
\begin{equation} \label{maxcoh}
\cohc := \left|\left\langle e^{i\phid({\xd},Y)} \right\rangle_{{\xd},Y}\right| = 
\sum_\info \wp(\info) \left|\ccc_{|\info} \right| ,
\end{equation}
with $\wp(Y)$ being the probability of obtaining the information $\info$. Thus $\left|\ccc_{|\info} \right|$ can be interpreted as conditional coherence. 

\section{SQ for noise sensing} 
As motivated in the introduction, a natural way to obtain  $\info$ is to use a SQ (s). This is similar to the data qubit but with a Hamiltonian $\frac{\ks}{2}  \hat\sigma_z^{\rm s}\rtp(t)$ making it more sensitive to the noise than the data qubit ($\ks \gg  \kd$).  
It can be frequently probed, as described by a projective measurement of the observable $\mathbb{1} - \ket{\theta}^{\rm s}\bra{\theta}$, yielding outcomes $y\in\{ 0,1\}$. Here $\ket{\theta}^{\rm s}$ is the equatorial state $(\ket{+1}^{\rm s}_z + e^{i\theta}\ket{-1}^{\rm s}_z)/\sqrt{2}$, with  $\hat\sigma_z^{\rm s}\ket{\pm1}^{\rm s}_z = \pm \ket{\pm1}^{\rm s}_z$. The SQ is initially prepared in the equatorial state $\ket{0}^{\rm s}$, and immediately reset to this after each  measurement. Between measurements it remains in an equatorial state, $\ket{\phis}^{\rm s}$, with $\dot\phis(t)=\ks \rtp(t)$. See Fig.~\ref{fig1}(a). 

Let us define $\{ t_1, t_2, \dots , t_N\equiv T\}$ as the times for SQ measurement and $y_n \in \{0,1\}$ as the corresponding measurement results. At any $t_n$, the SQ phase accumulated during the waiting time $\ddt_n := t_n - t_{n-1}$, between measurements, is
$\phis(t_n) = \ks x_n$, where $x_n :=  X(t_n)-X(t_{n-1})$ from \erf{IRTP}. We use Bayesian estimation and control based on the {\em likelihood function}
\be \label{eq:forwardP}
\wp(y_{n}|\theta_n,x_n) = \left| y_{n} - \cos^2\sq{\half (\theta_n - Kx_n ) } 
\right|, 
\ee
where $\theta_n$ is the equatorial measurement angle introduced above at time $t_n$.Now, given the record $Y_{n} := (y_1, y_2, \dots, y_{n})$, with $n\leq N$, the problem is to optimize the choice for the next waiting time $\ddt_{n+1}$ and measurement angle $\theta_{n+1}$ to maximize the reward function (\ref{maxcoh}). (Note that when  we condition on $Y_n$ we implicitly condition on all the previous choices of durations and angles which are functions of the earlier parts of $Y_n$.) This optimization requires us to better understand 
\erf{maxcoh}. 

\section{Bayesian maps for phase estimation} 
We show in the CP~\cite{PRA} that, from the dynamics of the RTP described in \erfand{MESS}{IRTP}, and the likelihood function (\ref{eq:forwardP}), we can apply Bayes' theorem to evaluate \erf{maxcoh} at $t=T$ as 
\be\label{costwithA}
\cohc(T) = 
\sum\nolimits_{Y_N} \left| \wp(Y_N) \ccc_{|Y_N} \right| = 
\sum\nolimits_{Y_N} \left| \underline{I}^\top\,  \underline{A}_N \right|.
\ee
Here $\underline{I} = \left( 1, 1 \right)\tp$, while  we define $\underline{A}_n$ for any $n$ as
\beq
\underline{A}_n = (A^{+1}_n,A^{-1}_n)\tp, \textrm{ where }
 A^{\rtp_n}_n := \wp(Y_n,\rtp_n)\ccc_{|Y_n,\rtp_n}, 
\nn
\eeq
and $\rtp_n:= \rtp(t_n)$. This can be efficiently calculated via   
\begin{equation} \label{Map1}
    \ul{A}_n = {\bf F}(\theta_n,\ddt_n, y_n)\,\cdots\, {\bf F}(\theta_1,\ddt_1, y_1)\,\underline{A}_0,
\end{equation}
where  each ${\bf F}(\theta,\ddt,y)$ is a $2\times 2$ complex matrix (one for each measurement), which also depends upon $\kd$, $\gu$, $\gd$, and $\ks$. Here, $\underline{A}_0=\ul{P}_0$ encodes the initial or prior probabilities for $\rtp_0$, which we take to be the unconditioned stationary probabilities $\ul{P}\ss$ of \erf{MESS}. The optimal final data qubit control in \erf{optctrl} can thus also be easily computed in an experiment as $c(Y_N) = \arg(\underline{I}^\top\, \underline{A}_N)$. The full expression for ${\bf F}$ is lengthy and is given in the \sm~\cite{PRA}. 

Recall that the task is to choose $\ddt_{n+1}$ and $\theta_{n+1}$ given $Y_{n}$ so as to maximize $\cohc(T)$. The significance of \erfand{costwithA}{Map1} is that the choice should depend only on $\ul{A}_n$~\cite{PRA}. Although $\ul{A}_n$, a complex 2-vector, encodes four real parameters, it can be shown~\cite{PRA} there are only two real {\em sufficient statistics} for the adaptive measurement choice, which we now define and explain.

If, hypothetically, we could find out $\rtp_n$, we could use that to refine the control that we would (hypothetically, if $t_n$ were $T$) apply to the data qubit, from $\cd_n(Y_n)=\arg \ccc_{|Y_n}$ to $\cd'_n(Y_n,z_n)=\arg \ccc_{|Y_n,z_n}$. The difference between the two refined values, $\cd'_n(Y_n,+1)-\cd'_n(Y_n,-1)$, scaled to be typically $O(1)$~\cite{PRA}, is  
\be  \label{defalpha}
\alpha_n := \frac{K}{\kappa}
\ro{\arg \ccc_{|Y_n,+1}-\arg \ccc_{|Y_n,-1} }
= \frac{K}{\kappa}\arg\frac{A_n^{z=+1}}{A_n^{z=-1}}.
\ee
This $\alpha$ is the first of the two sufficient statistics.

Now, in the limit $\kappa\ll K$, not only are the arguments of $\ccc_{|Y_n,\pm1}$ very close, as per \erf{defalpha}, but so are their moduli, with $|\ccc_{|Y_n,+1}|/|\ccc_{|Y_n,-1}| = 1 + O\big((\kappa/K)^2\big)$~\cite{PRA}. It follows, with relative errors of the same magnitude, that  $|A_n^{z_n}|\approx|\ccc_{|Y_n}|\, \wp(Y_n,z_n)$, and that 
\be \label{defzeta}
\zeta_n :=\ro{|A_n^{\rtp=+1}|-|A_n^{\rtp=-1}|} /\ro{|A_n^{\rtp=+1}|+|A_n^{\rtp=-1}|}
\ee
is approximately the mean of $\rtp_n$ conditioned on $Y_n$. This $\zeta$ is the second of the two sufficient statistics.  

\section{Greedy algorithm} 
The most obvious strategy, for choosing $\tau_{n+1}$ and $\theta_{n+1}$, is a `greedy' one. Given $Y_n$ but before $y_{n+1}$ is obtained, `Greedy' (as we call it) at time $t$ acts as if $T=t+\dd t$ (\ie\ as if the protocol were about to end) and maximizes the reward function at that time, $\cohc_{|Y_n}(t+\dd t)$, conditioned on the known information. This is explored in detail in the CP~\cite{PRA}. We find that, 
in the regime of \erf{AsymReg}, to an excellent approximation, Greedy chooses adaptively, as follows: 
\begin{align}  
\theta_{n+1} &= s_n \Theta_{\rm G}(\alpha_n,|\zeta_n|) 
\label{Greedyfunction}
\; \text{ where } s_n := {\rm sign}(\zeta_n), \\
\tau_{n+1} &= |\theta_{n+1}|/\ks. 
 \label{tauthetaconnect}
\end{align}
Here $\Theta_{\rm G} \approx \pi/2$, varying by only $2.5\%$ as its arguments vary. Since $s_n$ is the most likely value of $z_n$, 
the choice $\theta_{n+1} = \ks \tau_{n+1} s_n$ corresponds to measuring along the direction of the most likely state $\ket{\Phi}^{\rm s}$ of the SQ [Fig.~\ref{fig1}(c)], with $y=0$ being the most likely, or `null' result. Naively, one might expect $\Theta_{\rm G} = \pi/2$ to be optimal  since, ignoring jumps during the measurement interval, a duration of $\ddt = \pi/(2\ks)$ would map the two RTP states ($z_n=\pm 1$) to orthogonal states of the SQ ($\Phi=\pm \pi/2$). However, while Greedy does greatly suppress decoherence [Fig.~\ref{fig1}(b)], we can achieve more by optimizing $\Theta$, as we now show. 

\begin{figure}[t]
\includegraphics[width=0.95\columnwidth]{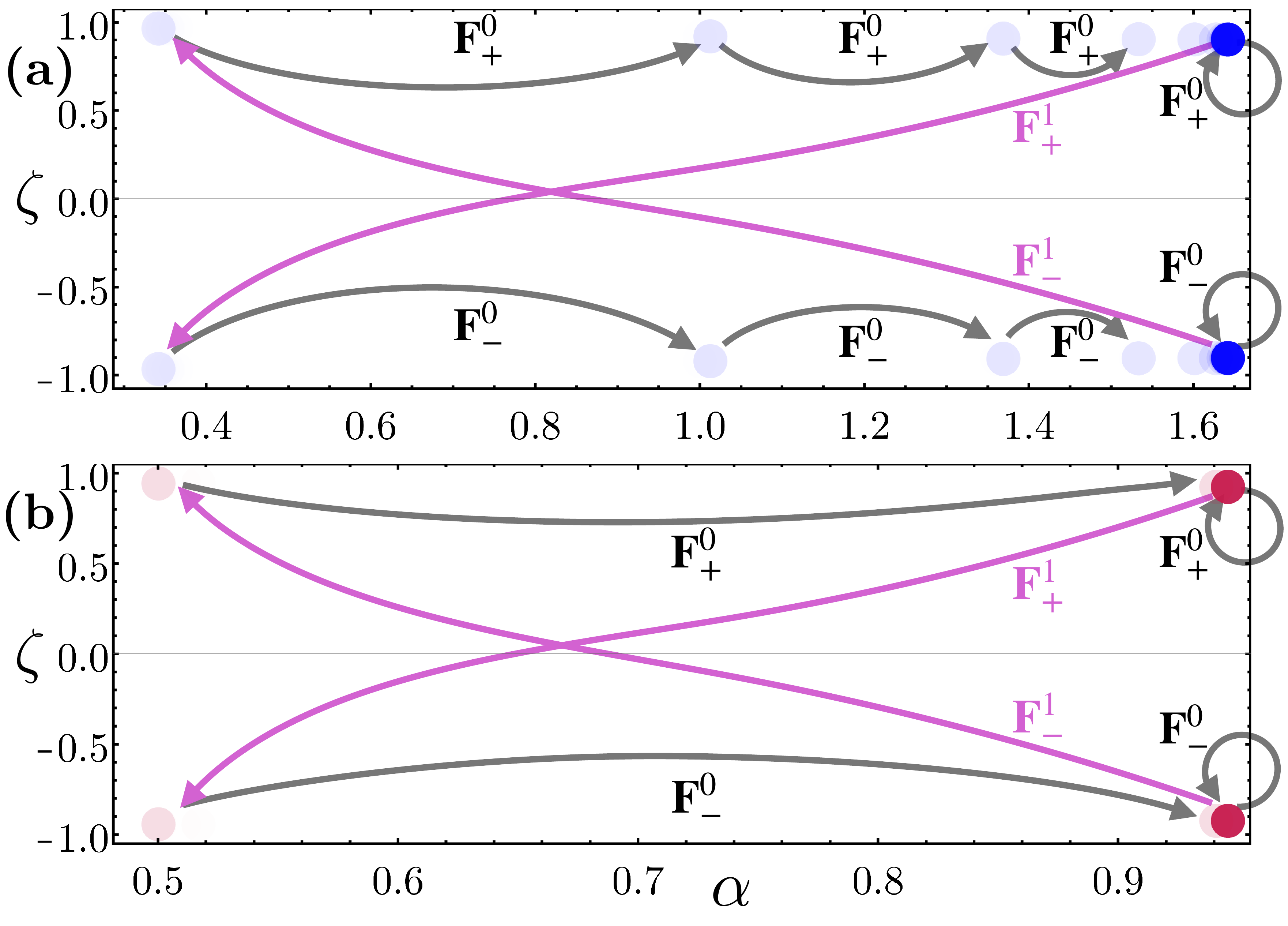}
\caption{\label{fig:phasespace} Phase-space ($\zeta,\alpha$) portrait showing jumps induced by measurement maps ${\bf F}_s^y$ for the constant-$\Theta$ adaptive scheme, 
with $\Theta=1.0$ (a) and $\Theta=\Theta\opt\approx1.50$ (b). Dots are all possible points after $n=10$ measurements with colour saturation showing relative probability. Parameters are as in Fig.~\ref{fig1}. 
}
\end{figure}

\section{Map-based Optimized Adaptive Algorithm for Asymptotic Regime (MOAAAR)} 
We use a map-based approach for our optimization, in order to obtain analytical results in the asymptotic regime (\ref{AsymReg}). We make the ans\"atze~(\ref{Greedyfunction}), (\ref{tauthetaconnect}) for the adaptive algorithm, but replace $\Theta_{\rm G}$ by $\Theta$, a constant. Hence there are only four possible values of the map ${\bf F}(\theta_n,\tau_n,y_n)$ in \erf{Map1}. These are ${\bf F}^y_s(\Theta) := {\bf F}(s\Theta,\Theta/K,y)$, with $y\in \{0,1\}$ and $s = \pm$. To understand and thus optimize our algorithm, we need to study the behaviour of the parameters $(\alpha,\zeta)$, which encode all our relevant knowledge.  

The behaviour is shown in Fig.~\ref{fig:phasespace}(a), choosing $\Theta = 1.0$ for  illustrative purposes. Under the mapping (\ref{Map1}), applied stochastically with the actual statistics of $Y_n$, we see that the `system' $(\alpha,\zeta)$ spends almost all of the time close to just two points. These are the fixed points (stable eigenstates) $\ul{E}_\pm^0$ of the maps ${\bf F}^0_\pm(\Theta)$ with the null outcome ($y=0$). Thus, if we ignore the initial [$t = O(1/\bg)$] and final [$T-t = O(1/\bg)$] transients, the dynamics of $(\alpha,\zeta)$ mirrors that of the RTP, with transition rates $\gu, \gd$. This is an ergodic process with steady state $P_{\rm ss}$ in \erf{MESS}. 

\begin{figure}[t]
\includegraphics[width=\columnwidth]{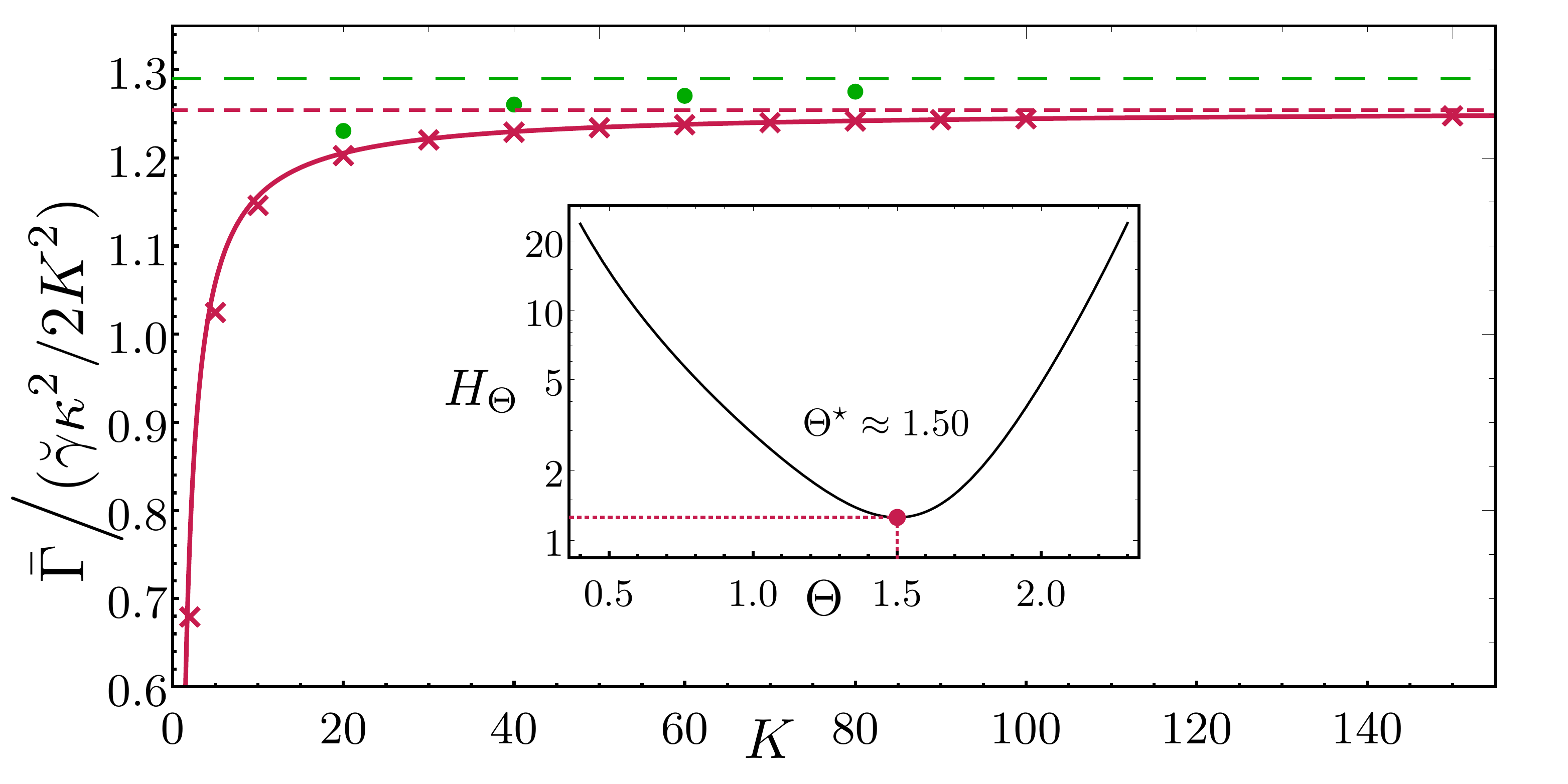}
\caption{\label{fig:decrat} Decoherence rate for Greedy (green dots) and MOAAAR (maroon crosses) versus $\ks$, scaled so as to asymptote to $O(1)$, with $\kd=0.2, \gu=\gd = 1$. Data points are from slope-fitting to exact numerical calculations of \erf{maxcoh}. Asymptotes (horizontal, dashed) are from \erf{defH} (plotted in the inset), with $\Theta=\pi/2$ (Greedy, green long dashes) and $\Theta=\Theta^\star$ (MOAAAR, maroon short dashes). Curve (solid) for MOAAAR is a closed-form approximation in the \sm~\cite{PRA}. 
}
\end{figure}

To maximize \erf{costwithA} for $T\gg \bg^{-1}$, transients can be ignored. Thus we seek a $\Theta$ that minimizes the decay of $\cohc(t)$ in the ergodic regime. Consider first the relative change, $\delta$, in the {\em conditional} coherence, $|\ccc_{|Y_n}|$. 
As the system state evolves from a known $\ul{A}_n$ at time $t_n$ to an unknown one at time $t_{n+1} = t_n + \tau$, this is given by 
\begin{align}
\delta(\ul{A}_n)  &:= \left[\left|\ccc_{|Y_n}\right|   -  \sum_{y_{n+1}} \wp(y_{n+1}|Y_n) \left|\ccc_{|Y_{n+1}}\right|\right] / \left|\ccc_{|Y_n}\right| \nn \\
&= 1 - | \underline{I} ^\top \underline{A}_n |^{-1} \sum\nolimits_{y} | \underline{I} ^\top\, {\bf F}_{s_n}^y\!(\Theta) \,\underline{A}_n|.
\end{align}
From the above considerations of ergodicity, it can be shown~\cite{PRA} that the decay {\em rate} $\bar\Gamma$ of $\cohc$ can be evaluated as  $\bar\delta/\tau$, where the bar indicates the steady-state ensemble average. That is, emphasising the $\Theta$-dependence, 
\beq \label{eq:asymav}
\bar\Gamma(\Theta) := \sum_{s=\pm} {P}_{\rm ss}(s)
\frac{ | \underline{I}^\top \underline{E}_s^0 | - \sum_{y} | \underline{I} ^\top\, {\bf F}_s^y(\Theta) \, \underline{E}_s^0 |  }{\ddt | \underline{I} ^\top \underline{E}_s^0 |}.
\eeq
Finally, a lengthy calculation~\cite{PRA} reveals that in the regime (\ref{AsymReg}), 
$\bar\Gamma(\Theta)  \to H_{\Theta}  \bg \kd^2/(2\ks^2)$, 
where $H_{\Theta}$ equals 
\begin{align}
 3\Theta^2 \csc^4 \Theta - [2\Theta (\Theta - \cot\Theta) +1] {\rm csc}^2\Theta + \tfrac{1}{3}\Theta^2  - 1.  \label{defH}
\end{align}
This function is plotted in the inset of Fig.~\ref{fig:decrat}. 
Its minimum is $H\opt \approx 1.254$ at $\Theta = \Theta\opt \approx 1.50055$. This optimization defines our MOAAAR protocol, the one giving the minimum decoherence rate of 
$\bar\Gamma(\Theta\opt) = H\opt {\bg \kd^2}/(2\ks^2).$
Comparing with the no-control case (\ref{uncondec}), gives the earlier quoted quadratic decoherence suppression, Eq.~\eqref{multfac}. 

We show the dynamics of phase space ($\alpha, \zeta$) for $\Theta=\Theta\opt$ in  Fig.~\ref{fig:phasespace}(b). As expected, the system spends almost all its time in the two fixed points $E^0_\pm$, until a non-null result ($y=1$) causes a jump to $ {\bf F}^1_\pm E^0_\pm$. But, unlike in Fig.~\ref{fig:phasespace}(a) ($\Theta = 1.0$), at the next measurement the system almost always jumps practically the whole way back to a fixed point $E^0_\mp$ (the opposite one whence it started). Thus the entire evolution is, for all practical purposes, confined to four points, $E^0_\pm$ and ${\bf F}^1_\pm E^0_\pm$.

The fact that $\Theta\opt$ is significantly different from $\pi/2$ (as chosen by Greedy almost all the time in the asymptotic limit), highlights the nontriviality of MOAAAR. It also means that to 
 implement MOAAAR experimentally would require a real-time feedback loop,   
albeit quite a simple one: one simply switches the sign of $\theta$ whenever a non-null result ($y=1$) is obtained. 
In Fig.~\ref{fig:decrat} we confirm by exact numerics that MOAAAR  outperforms Greedy  and that their decoherence rates, scaled by $2\ks^2/(\bg\kd^2)$, asymptote to $H^\star$ and $H_{\pi/2} \approx 1.290$ respectively. In the \sm~\cite{PRA} we provide further evidence supporting the plausibility of the optimality of MOAAAR, by showing numerically that it is also the optimal algorithm out of a larger family of algorithms. Specifically, we optimize over two parameters, $\Theta$ and $\tau$, where $\tau$ is fixed but not constrained by $\tau = \Theta/K$. 

\section{Discussion}  For suppression of RTP phase noise, using information obtained from a SQ, we have proposed a protocol (MOAAAR) --- an adaptive sequence of projective measurements on the SQ, followed by a control on the data qubit at the final time --- which is  plausibly optimal in the good parameter regime. In this regime, the suppression of the decoherence rate, \erf{multfac}, is limited only by the SQ sensitivity. A conceptually simpler but implementationally more complicated algorithm, `Greedy', performs slightly worse. In the \sm~\cite{PRA} we perform further comparisons of these two strategies. 

This work establishes that, like the well-known noise-mitigation strategies of DD and QEC, the SQ paradigm can work arbitrarily well in a suitable regime. The approach we use here can certainly be generalized for more complicated noise processes, such as multiple RTPs or multi-level RTPs. This, and other directions to study the applicability of SQs in real-world situations, are discussed in the \sm~\cite{PRA}. 

\begin{acknowledgments} \vspace{2ex}
This work was supported by the Australian Government via the Australia-US-MURI grant
AUSMURI000002, by the Australian Research Council via the Centre of Excellence grant CE170100012, and by National Research Council of Thailand (NRCT) grant N41A640120. A.C.~also acknowledges the support from the NSRF via the Program Management Unit for Human Resources and Institutional Development, Research and Innovation, grant number B05F650024. We thank members of the AUSMURI collaboration, especially Gerardo Paz Silva, Andrea Morello, and Ken Brown, for feedback on this work. We acknowledge the Yuggera people, the traditional owners of the land at at Griffith University on which this work was undertaken. 

H.S. and A.C. contributed equally to this work.
\end{acknowledgments}

\bibliographystyle{apsrev4-1}

\begin{thebibliography}{39}%
\makeatletter
\providecommand \@ifxundefined [1]{%
 \@ifx{#1\undefined}
}%
\providecommand \@ifnum [1]{%
 \ifnum #1\expandafter \@firstoftwo
 \else \expandafter \@secondoftwo
 \fi
}%
\providecommand \@ifx [1]{%
 \ifx #1\expandafter \@firstoftwo
 \else \expandafter \@secondoftwo
 \fi
}%
\providecommand \natexlab [1]{#1}%
\providecommand \enquote  [1]{``#1''}%
\providecommand \bibnamefont  [1]{#1}%
\providecommand \bibfnamefont [1]{#1}%
\providecommand \citenamefont [1]{#1}%
\providecommand \href@noop [0]{\@secondoftwo}%
\providecommand \href [0]{\begingroup \@sanitize@url \@href}%
\providecommand \@href[1]{\@@startlink{#1}\@@href}%
\providecommand \@@href[1]{\endgroup#1\@@endlink}%
\providecommand \@sanitize@url [0]{\catcode `\\12\catcode `\$12\catcode
  `\&12\catcode `\#12\catcode `\^12\catcode `\_12\catcode `\%12\relax}%
\providecommand \@@startlink[1]{}%
\providecommand \@@endlink[0]{}%
\providecommand \url  [0]{\begingroup\@sanitize@url \@url }%
\providecommand \@url [1]{\endgroup\@href {#1}{\urlprefix }}%
\providecommand \urlprefix  [0]{URL }%
\providecommand \Eprint [0]{\href }%
\providecommand \doibase [0]{https://doi.org/}%
\providecommand \selectlanguage [0]{\@gobble}%
\providecommand \bibinfo  [0]{\@secondoftwo}%
\providecommand \bibfield  [0]{\@secondoftwo}%
\providecommand \translation [1]{[#1]}%
\providecommand \BibitemOpen [0]{}%
\providecommand \bibitemStop [0]{}%
\providecommand \bibitemNoStop [0]{.\EOS\space}%
\providecommand \EOS [0]{\spacefactor3000\relax}%
\providecommand \BibitemShut  [1]{\csname bibitem#1\endcsname}%
\let\auto@bib@innerbib\@empty
\bibitem [{\citenamefont {Arute}\ \emph {et~al.}(2019)\citenamefont {Arute},
  \citenamefont {Arya}, \citenamefont {Babbush} \emph {et~al.}}]{AruAry2019}%
  \BibitemOpen
  \bibfield  {author} {\bibinfo {author} {\bibfnamefont {F.}~\bibnamefont
  {Arute}}, \bibinfo {author} {\bibfnamefont {K.}~\bibnamefont {Arya}},
  \bibinfo {author} {\bibfnamefont {R.}~\bibnamefont {Babbush}}, \emph
  {et~al.},\ }\href {https://doi.org/10.1038/s41586-019-1666-5} {\bibfield
  {journal} {\bibinfo  {journal} {Nature}\ }\textbf {\bibinfo {volume} {574}},\
  \bibinfo {pages} {505} (\bibinfo {year} {2019})}\BibitemShut {NoStop}%
\bibitem [{\citenamefont {Zhong}\ \emph {et~al.}(2020)\citenamefont {Zhong},
  \citenamefont {Wang}, \citenamefont {Deng} \emph {et~al.}}]{ZhoWan2020}%
  \BibitemOpen
  \bibfield  {author} {\bibinfo {author} {\bibfnamefont {H.-S.}\ \bibnamefont
  {Zhong}}, \bibinfo {author} {\bibfnamefont {H.}~\bibnamefont {Wang}},
  \bibinfo {author} {\bibfnamefont {Y.-H.}\ \bibnamefont {Deng}}, \emph
  {et~al.},\ }\href {https://doi.org/10.1126/science.abe8770} {\bibfield
  {journal} {\bibinfo  {journal} {Science}\ }\textbf {\bibinfo {volume}
  {370}},\ \bibinfo {pages} {1460} (\bibinfo {year} {2020})}\BibitemShut
  {NoStop}%
\bibitem [{\citenamefont {Temme}\ \emph {et~al.}(2017)\citenamefont {Temme},
  \citenamefont {Bravyi},\ and\ \citenamefont {Gambetta}}]{TemBra2017}%
  \BibitemOpen
  \bibfield  {author} {\bibinfo {author} {\bibfnamefont {K.}~\bibnamefont
  {Temme}}, \bibinfo {author} {\bibfnamefont {S.}~\bibnamefont {Bravyi}},\ and\
  \bibinfo {author} {\bibfnamefont {J.~M.}\ \bibnamefont {Gambetta}},\ }\href
  {https://doi.org/10.1103/PhysRevLett.119.180509} {\bibfield  {journal}
  {\bibinfo  {journal} {Phys. Rev. Lett.}\ }\textbf {\bibinfo {volume} {119}},\
  \bibinfo {pages} {180509} (\bibinfo {year} {2017})}\BibitemShut {NoStop}%
\bibitem [{\citenamefont {Preskill}(2018)}]{Pre2018}%
  \BibitemOpen
  \bibfield  {author} {\bibinfo {author} {\bibfnamefont {J.}~\bibnamefont
  {Preskill}},\ }\href {https://doi.org/10.22331/q-2018-08-06-79} {\bibfield
  {journal} {\bibinfo  {journal} {{Quantum}}\ }\textbf {\bibinfo {volume}
  {2}},\ \bibinfo {pages} {79} (\bibinfo {year} {2018})}\BibitemShut {NoStop}%
\bibitem [{\citenamefont {Endo}\ \emph {et~al.}(2018)\citenamefont {Endo},
  \citenamefont {Benjamin},\ and\ \citenamefont {Li}}]{EndBen2018}%
  \BibitemOpen
  \bibfield  {author} {\bibinfo {author} {\bibfnamefont {S.}~\bibnamefont
  {Endo}}, \bibinfo {author} {\bibfnamefont {S.~C.}\ \bibnamefont {Benjamin}},\
  and\ \bibinfo {author} {\bibfnamefont {Y.}~\bibnamefont {Li}},\ }\href
  {https://doi.org/10.1103/PhysRevX.8.031027} {\bibfield  {journal} {\bibinfo
  {journal} {Phys. Rev. X}\ }\textbf {\bibinfo {volume} {8}},\ \bibinfo {pages}
  {031027} (\bibinfo {year} {2018})}\BibitemShut {NoStop}%
\bibitem [{\citenamefont {Kandala}\ \emph {et~al.}(2019)\citenamefont
  {Kandala}, \citenamefont {Temme}, \citenamefont {C{\'{o}}rcoles},
  \citenamefont {Mezzacapo}, \citenamefont {Chow},\ and\ \citenamefont
  {Gambetta}}]{KanTem2019}%
  \BibitemOpen
  \bibfield  {author} {\bibinfo {author} {\bibfnamefont {A.}~\bibnamefont
  {Kandala}}, \bibinfo {author} {\bibfnamefont {K.}~\bibnamefont {Temme}},
  \bibinfo {author} {\bibfnamefont {A.~D.}\ \bibnamefont {C{\'{o}}rcoles}},
  \bibinfo {author} {\bibfnamefont {A.}~\bibnamefont {Mezzacapo}}, \bibinfo
  {author} {\bibfnamefont {J.~M.}\ \bibnamefont {Chow}},\ and\ \bibinfo
  {author} {\bibfnamefont {J.~M.}\ \bibnamefont {Gambetta}},\ }\href
  {https://doi.org/10.1038/s41586-019-1040-7} {\bibfield  {journal} {\bibinfo
  {journal} {Nature}\ }\textbf {\bibinfo {volume} {567}},\ \bibinfo {pages}
  {491} (\bibinfo {year} {2019})}\BibitemShut {NoStop}%
\bibitem [{\citenamefont {Viola}\ \emph {et~al.}(1999)\citenamefont {Viola},
  \citenamefont {Knill},\ and\ \citenamefont {Lloyd}}]{Vio99}%
  \BibitemOpen
  \bibfield  {author} {\bibinfo {author} {\bibfnamefont {L.}~\bibnamefont
  {Viola}}, \bibinfo {author} {\bibfnamefont {E.}~\bibnamefont {Knill}},\ and\
  \bibinfo {author} {\bibfnamefont {S.}~\bibnamefont {Lloyd}},\ }\href
  {https://doi.org/10.1103/PhysRevLett.82.2417} {\bibfield  {journal} {\bibinfo
   {journal} {Phys. Rev. Lett.}\ }\textbf {\bibinfo {volume} {82}},\ \bibinfo
  {pages} {2417} (\bibinfo {year} {1999})}\BibitemShut {NoStop}%
\bibitem [{\citenamefont {Viola}\ and\ \citenamefont
  {Knill}(2003)}]{viola2003robust}%
  \BibitemOpen
  \bibfield  {author} {\bibinfo {author} {\bibfnamefont {L.}~\bibnamefont
  {Viola}}\ and\ \bibinfo {author} {\bibfnamefont {E.}~\bibnamefont {Knill}},\
  }\href {https://doi.org/10.1103/PhysRevLett.90.037901} {\bibfield  {journal}
  {\bibinfo  {journal} {Phys. Rev. Lett.}\ }\textbf {\bibinfo {volume} {90}},\
  \bibinfo {pages} {037901} (\bibinfo {year} {2003})}\BibitemShut {NoStop}%
\bibitem [{\citenamefont {Biercuk}\ \emph {et~al.}(2011)\citenamefont
  {Biercuk}, \citenamefont {Doherty},\ and\ \citenamefont
  {Uys}}]{biercuk2011dynamical}%
  \BibitemOpen
  \bibfield  {author} {\bibinfo {author} {\bibfnamefont {M.}~\bibnamefont
  {Biercuk}}, \bibinfo {author} {\bibfnamefont {A.}~\bibnamefont {Doherty}},\
  and\ \bibinfo {author} {\bibfnamefont {H.}~\bibnamefont {Uys}},\ }\href
  {https://doi.org/10.1088/0953-4075/44/15/154002} {\bibfield  {journal}
  {\bibinfo  {journal} {Journal of Physics B: Atomic, Molecular and Optical
  Physics}\ }\textbf {\bibinfo {volume} {44}},\ \bibinfo {pages} {154002}
  (\bibinfo {year} {2011})}\BibitemShut {NoStop}%
\bibitem [{\citenamefont {Ng}\ \emph {et~al.}(2011)\citenamefont {Ng},
  \citenamefont {Lidar},\ and\ \citenamefont {Preskill}}]{NgLid2011}%
  \BibitemOpen
  \bibfield  {author} {\bibinfo {author} {\bibfnamefont {H.~K.}\ \bibnamefont
  {Ng}}, \bibinfo {author} {\bibfnamefont {D.~A.}\ \bibnamefont {Lidar}},\ and\
  \bibinfo {author} {\bibfnamefont {J.}~\bibnamefont {Preskill}},\ }\href
  {https://doi.org/10.1103/PhysRevA.84.012305} {\bibfield  {journal} {\bibinfo
  {journal} {Phys. Rev. A}\ }\textbf {\bibinfo {volume} {84}},\ \bibinfo
  {pages} {012305} (\bibinfo {year} {2011})}\BibitemShut {NoStop}%
\bibitem [{\citenamefont {Souza}\ \emph {et~al.}(2011)\citenamefont {Souza},
  \citenamefont {\'Alvarez},\ and\ \citenamefont {Suter}}]{SouAlv2011}%
  \BibitemOpen
  \bibfield  {author} {\bibinfo {author} {\bibfnamefont {A.~M.}\ \bibnamefont
  {Souza}}, \bibinfo {author} {\bibfnamefont {G.~A.}\ \bibnamefont
  {\'Alvarez}},\ and\ \bibinfo {author} {\bibfnamefont {D.}~\bibnamefont
  {Suter}},\ }\href {https://doi.org/10.1103/PhysRevLett.106.240501} {\bibfield
   {journal} {\bibinfo  {journal} {Phys. Rev. Lett.}\ }\textbf {\bibinfo
  {volume} {106}},\ \bibinfo {pages} {240501} (\bibinfo {year}
  {2011})}\BibitemShut {NoStop}%
\bibitem [{\citenamefont {Medford}\ \emph {et~al.}(2012)\citenamefont
  {Medford}, \citenamefont {Cywi\ifmmode~\acute{n}\else \'{n}\fi{}ski},
  \citenamefont {Barthel}, \citenamefont {Marcus}, \citenamefont {Hanson},\
  and\ \citenamefont {Gossard}}]{MedCyw2012}%
  \BibitemOpen
  \bibfield  {author} {\bibinfo {author} {\bibfnamefont {J.}~\bibnamefont
  {Medford}}, \bibinfo {author} {\bibfnamefont {L.}~\bibnamefont
  {Cywi\ifmmode~\acute{n}\else \'{n}\fi{}ski}}, \bibinfo {author}
  {\bibfnamefont {C.}~\bibnamefont {Barthel}}, \bibinfo {author} {\bibfnamefont
  {C.~M.}\ \bibnamefont {Marcus}}, \bibinfo {author} {\bibfnamefont {M.~P.}\
  \bibnamefont {Hanson}},\ and\ \bibinfo {author} {\bibfnamefont {A.~C.}\
  \bibnamefont {Gossard}},\ }\href
  {https://doi.org/10.1103/PhysRevLett.108.086802} {\bibfield  {journal}
  {\bibinfo  {journal} {Phys. Rev. Lett.}\ }\textbf {\bibinfo {volume} {108}},\
  \bibinfo {pages} {086802} (\bibinfo {year} {2012})}\BibitemShut {NoStop}%
\bibitem [{\citenamefont {Paz-Silva}\ and\ \citenamefont
  {Lidar}(2013)}]{PazLid2013}%
  \BibitemOpen
  \bibfield  {author} {\bibinfo {author} {\bibfnamefont {G.~A.}\ \bibnamefont
  {Paz-Silva}}\ and\ \bibinfo {author} {\bibfnamefont {D.~A.}\ \bibnamefont
  {Lidar}},\ }\href {https://doi.org/10.1038/srep01530} {\bibfield  {journal}
  {\bibinfo  {journal} {Scientific Reports}\ }\textbf {\bibinfo {volume} {3}},\
  \bibinfo {pages} {1530} (\bibinfo {year} {2013})}\BibitemShut {NoStop}%
\bibitem [{\citenamefont {Zhang}\ \emph {et~al.}(2014)\citenamefont {Zhang},
  \citenamefont {Souza}, \citenamefont {Brandao},\ and\ \citenamefont
  {Suter}}]{ZhaSou2014}%
  \BibitemOpen
  \bibfield  {author} {\bibinfo {author} {\bibfnamefont {J.}~\bibnamefont
  {Zhang}}, \bibinfo {author} {\bibfnamefont {A.~M.}\ \bibnamefont {Souza}},
  \bibinfo {author} {\bibfnamefont {F.~D.}\ \bibnamefont {Brandao}},\ and\
  \bibinfo {author} {\bibfnamefont {D.}~\bibnamefont {Suter}},\ }\href
  {https://doi.org/10.1103/PhysRevLett.112.050502} {\bibfield  {journal}
  {\bibinfo  {journal} {Phys. Rev. Lett.}\ }\textbf {\bibinfo {volume} {112}},\
  \bibinfo {pages} {050502} (\bibinfo {year} {2014})}\BibitemShut {NoStop}%
\bibitem [{\citenamefont {Shor}(1995)}]{Shor1995}%
  \BibitemOpen
  \bibfield  {author} {\bibinfo {author} {\bibfnamefont {P.~W.}\ \bibnamefont
  {Shor}},\ }\href {https://doi.org/10.1103/PhysRevA.52.R2493} {\bibfield
  {journal} {\bibinfo  {journal} {Phys. Rev. A}\ }\textbf {\bibinfo {volume}
  {52}},\ \bibinfo {pages} {R2493} (\bibinfo {year} {1995})}\BibitemShut
  {NoStop}%
\bibitem [{\citenamefont {Steane}(1996)}]{Steane1996}%
  \BibitemOpen
  \bibfield  {author} {\bibinfo {author} {\bibfnamefont {A.~M.}\ \bibnamefont
  {Steane}},\ }\href {https://doi.org/10.1103/PhysRevLett.77.793} {\bibfield
  {journal} {\bibinfo  {journal} {Phys. Rev. Lett.}\ }\textbf {\bibinfo
  {volume} {77}},\ \bibinfo {pages} {793} (\bibinfo {year} {1996})}\BibitemShut
  {NoStop}%
\bibitem [{\citenamefont {Terhal}(2015)}]{Terhal2015}%
  \BibitemOpen
  \bibfield  {author} {\bibinfo {author} {\bibfnamefont {B.~M.}\ \bibnamefont
  {Terhal}},\ }\href {https://doi.org/10.1103/RevModPhys.87.307} {\bibfield
  {journal} {\bibinfo  {journal} {Rev. Mod. Phys.}\ }\textbf {\bibinfo {volume}
  {87}},\ \bibinfo {pages} {307} (\bibinfo {year} {2015})}\BibitemShut
  {NoStop}%
\bibitem [{\citenamefont {Gupta}\ \emph {et~al.}(2020)\citenamefont {Gupta},
  \citenamefont {Edmunds}, \citenamefont {Milne}, \citenamefont {Hempel},\ and\
  \citenamefont {Biercuk}}]{GupEdm2020}%
  \BibitemOpen
  \bibfield  {author} {\bibinfo {author} {\bibfnamefont {R.~S.}\ \bibnamefont
  {Gupta}}, \bibinfo {author} {\bibfnamefont {C.~L.}\ \bibnamefont {Edmunds}},
  \bibinfo {author} {\bibfnamefont {A.~R.}\ \bibnamefont {Milne}}, \bibinfo
  {author} {\bibfnamefont {C.}~\bibnamefont {Hempel}},\ and\ \bibinfo {author}
  {\bibfnamefont {M.~J.}\ \bibnamefont {Biercuk}},\ }\href
  {https://doi.org/10.1038/s41534-020-0286-0} {\bibfield  {journal} {\bibinfo
  {journal} {npj Quantum Information}\ }\textbf {\bibinfo {volume} {6}},\
  \bibinfo {pages} {53} (\bibinfo {year} {2020})}\BibitemShut {NoStop}%
\bibitem [{\citenamefont {Majumder}\ \emph {et~al.}(2020)\citenamefont
  {Majumder}, \citenamefont {{Andreta de Castro}},\ and\ \citenamefont
  {Brown}}]{MajAnd2020}%
  \BibitemOpen
  \bibfield  {author} {\bibinfo {author} {\bibfnamefont {S.}~\bibnamefont
  {Majumder}}, \bibinfo {author} {\bibfnamefont {L.}~\bibnamefont {{Andreta de
  Castro}}},\ and\ \bibinfo {author} {\bibfnamefont {K.~R.}\ \bibnamefont
  {Brown}},\ }\href {https://doi.org/10.1038/s41534-020-0251-y} {\bibfield
  {journal} {\bibinfo  {journal} {npj Quantum Information}\ }\textbf {\bibinfo
  {volume} {6}},\ \bibinfo {pages} {19} (\bibinfo {year} {2020})}\BibitemShut
  {NoStop}%
\bibitem [{\citenamefont {Singh}\ \emph {et~al.}(2022)\citenamefont {Singh},
  \citenamefont {Conor E.~Bradley}, \citenamefont {Ramesh}, \citenamefont
  {White},\ and\ \citenamefont {Bernien}}]{SinBra2022}%
  \BibitemOpen
  \bibfield  {author} {\bibinfo {author} {\bibfnamefont {K.}~\bibnamefont
  {Singh}}, \bibinfo {author} {\bibfnamefont {C.~E.}\ \bibnamefont {Bradley}}, \bibinfo {author} {\bibfnamefont {S.}~\bibnamefont
  {Anand}}, \bibinfo {author} {\bibfnamefont {V.}~\bibnamefont {Ramesh}},
  \bibinfo {author} {\bibfnamefont {R.}~\bibnamefont {White}},\ and\ \bibinfo
  {author} {\bibfnamefont {H.}~\bibnamefont {Bernien}},\ }\href@noop {}
  {\bibfield  {journal} {\bibinfo  {journal} {arXiv:2208.11716 [quant-ph]}\ }
  (\bibinfo {year} {2022})}\BibitemShut {NoStop}%
\bibitem [{\citenamefont {Itakura}\ and\ \citenamefont
  {Tokura}(2003)}]{ItaTok2003}%
  \BibitemOpen
  \bibfield  {author} {\bibinfo {author} {\bibfnamefont {T.}~\bibnamefont
  {Itakura}}\ and\ \bibinfo {author} {\bibfnamefont {Y.}~\bibnamefont
  {Tokura}},\ }\href {https://doi.org/10.1103/PhysRevB.67.195320} {\bibfield
  {journal} {\bibinfo  {journal} {Phys. Rev. B}\ }\textbf {\bibinfo {volume}
  {67}},\ \bibinfo {pages} {195320} (\bibinfo {year} {2003})}\BibitemShut
  {NoStop}%
\bibitem [{\citenamefont {Galperin}\ \emph {et~al.}(2006)\citenamefont
  {Galperin}, \citenamefont {Altshuler}, \citenamefont {Bergli},\ and\
  \citenamefont {Shantsev}}]{GalAlt2006}%
  \BibitemOpen
  \bibfield  {author} {\bibinfo {author} {\bibfnamefont {Y.~M.}\ \bibnamefont
  {Galperin}}, \bibinfo {author} {\bibfnamefont {B.~L.}\ \bibnamefont
  {Altshuler}}, \bibinfo {author} {\bibfnamefont {J.}~\bibnamefont {Bergli}},\
  and\ \bibinfo {author} {\bibfnamefont {D.~V.}\ \bibnamefont {Shantsev}},\
  }\href {https://doi.org/10.1103/PhysRevLett.96.097009} {\bibfield  {journal}
  {\bibinfo  {journal} {Phys. Rev. Lett.}\ }\textbf {\bibinfo {volume} {96}},\
  \bibinfo {pages} {097009} (\bibinfo {year} {2006})}\BibitemShut {NoStop}%
\bibitem [{\citenamefont {Culcer}\ \emph {et~al.}(2009)\citenamefont {Culcer},
  \citenamefont {Hu},\ and\ \citenamefont {Das~Sarma}}]{CulHu2009}%
  \BibitemOpen
  \bibfield  {author} {\bibinfo {author} {\bibfnamefont {D.}~\bibnamefont
  {Culcer}}, \bibinfo {author} {\bibfnamefont {X.}~\bibnamefont {Hu}},\ and\
  \bibinfo {author} {\bibfnamefont {S.}~\bibnamefont {Das~Sarma}},\ }\href
  {https://doi.org/10.1063/1.3194778} {\bibfield  {journal} {\bibinfo
  {journal} {Applied Physics Letters}\ }\textbf {\bibinfo {volume} {95}},\
  \bibinfo {pages} {073102} (\bibinfo {year} {2009})}\BibitemShut {NoStop}%
\bibitem [{\citenamefont {Bergli}\ \emph {et~al.}(2009)\citenamefont {Bergli},
  \citenamefont {Galperin},\ and\ \citenamefont {Altshuler}}]{BerGal2009}%
  \BibitemOpen
  \bibfield  {author} {\bibinfo {author} {\bibfnamefont {J.}~\bibnamefont
  {Bergli}}, \bibinfo {author} {\bibfnamefont {Y.~M.}\ \bibnamefont
  {Galperin}},\ and\ \bibinfo {author} {\bibfnamefont {B.~L.}\ \bibnamefont
  {Altshuler}},\ }\href {https://doi.org/10.1088/1367-2630/11/2/025002}
  {\bibfield  {journal} {\bibinfo  {journal} {New Journal of Physics}\ }\textbf
  {\bibinfo {volume} {11}},\ \bibinfo {pages} {025002} (\bibinfo {year}
  {2009})}\BibitemShut {NoStop}%
\bibitem [{\citenamefont {Helstrom}(1976)}]{Helstrom}%
  \BibitemOpen
  \bibfield  {author} {\bibinfo {author} {\bibfnamefont {C.~W.}\ \bibnamefont
  {Helstrom}},\ }\href@noop {} {\emph {\bibinfo {title} {Quantum Detection and
  Estimation Theory}}},\ \bibinfo {series} {Mathematics in Science and
  Engineering}, Vol.\ \bibinfo {volume} {123}\ (\bibinfo  {publisher} {Academic
  Press},\ \bibinfo {address} {New York},\ \bibinfo {year} {1976})\BibitemShut
  {NoStop}%
\bibitem [{\citenamefont {Wiseman}\ and\ \citenamefont
  {Milburn}(2010)}]{WisMil10}%
  \BibitemOpen
  \bibfield  {author} {\bibinfo {author} {\bibfnamefont {H.~M.}\ \bibnamefont
  {Wiseman}}\ and\ \bibinfo {author} {\bibfnamefont {G.~J.}\ \bibnamefont
  {Milburn}},\ }\href@noop {} {\emph {\bibinfo {title} {Quantum Measurement and
  Control}}}\ (\bibinfo  {publisher} {Cambridge University Press},\ \bibinfo
  {address} {Cambridge, England},\ \bibinfo {year} {2010})\BibitemShut
  {NoStop}%
\bibitem [{\citenamefont {{Ac\'in}}\ \emph {et~al.}(2005)\citenamefont
  {{Ac\'in}}, \citenamefont {Bagan}, \citenamefont {Baig}, \citenamefont
  {\protect{Ll.} Masanes},\ and\ \citenamefont {{Mu\~noz-Tapia}}}]{Acin2005}%
  \BibitemOpen
  \bibfield  {author} {\bibinfo {author} {\bibfnamefont {A.}~\bibnamefont
  {{Ac\'in}}}, \bibinfo {author} {\bibfnamefont {E.}~\bibnamefont {Bagan}},
  \bibinfo {author} {\bibfnamefont {M.}~\bibnamefont {Baig}}, \bibinfo {author}
  {\bibnamefont {\protect{Ll.} Masanes}},\ and\ \bibinfo {author}
  {\bibfnamefont {R.}~\bibnamefont {{Mu\~noz-Tapia}}},\ }\href@noop {}
  {\bibfield  {journal} {\bibinfo  {journal} {Phys. Rev. A}\ }\textbf {\bibinfo
  {volume} {71}},\ \bibinfo {pages} {032338} (\bibinfo {year}
  {2005})}\BibitemShut {NoStop}%
\bibitem [{\citenamefont {Higgins}\ \emph {et~al.}(2009)\citenamefont
  {Higgins}, \citenamefont {Booth}, \citenamefont {Doherty}, \citenamefont
  {Bartlett}, \citenamefont {Wiseman},\ and\ \citenamefont
  {Pryde}}]{Higgins-discrim09}%
  \BibitemOpen
  \bibfield  {author} {\bibinfo {author} {\bibfnamefont {B.~L.}\ \bibnamefont
  {Higgins}}, \bibinfo {author} {\bibfnamefont {B.~M.}\ \bibnamefont {Booth}},
  \bibinfo {author} {\bibfnamefont {A.~C.}\ \bibnamefont {Doherty}}, \bibinfo
  {author} {\bibfnamefont {S.~D.}\ \bibnamefont {Bartlett}}, \bibinfo {author}
  {\bibfnamefont {H.~M.}\ \bibnamefont {Wiseman}},\ and\ \bibinfo {author}
  {\bibfnamefont {G.~J.}\ \bibnamefont {Pryde}},\ }\href
  {https://doi.org/10.1103/PhysRevLett.103.220503} {\bibfield  {journal}
  {\bibinfo  {journal} {Phys. Rev. Lett.}\ }\textbf {\bibinfo {volume} {103}},\
  \bibinfo {pages} {220503} (\bibinfo {year} {2009})}\BibitemShut {NoStop}%
\bibitem [{\citenamefont {Higgins}\ \emph {et~al.}(2011)\citenamefont
  {Higgins}, \citenamefont {Doherty}, \citenamefont {Bartlett}, \citenamefont
  {Pryde},\ and\ \citenamefont {Wiseman}}]{Higgins11}%
  \BibitemOpen
  \bibfield  {author} {\bibinfo {author} {\bibfnamefont {B.~L.}\ \bibnamefont
  {Higgins}}, \bibinfo {author} {\bibfnamefont {A.~C.}\ \bibnamefont
  {Doherty}}, \bibinfo {author} {\bibfnamefont {S.~D.}\ \bibnamefont
  {Bartlett}}, \bibinfo {author} {\bibfnamefont {G.~J.}\ \bibnamefont
  {Pryde}},\ and\ \bibinfo {author} {\bibfnamefont {H.~M.}\ \bibnamefont
  {Wiseman}},\ }\href {https://doi.org/10.1103/PhysRevA.83.052314} {\bibfield
  {journal} {\bibinfo  {journal} {Phys. Rev. A}\ }\textbf {\bibinfo {volume}
  {83}},\ \bibinfo {pages} {052314} (\bibinfo {year} {2011})}\BibitemShut
  {NoStop}%
\bibitem [{\citenamefont {Slussarenko}\ \emph {et~al.}(2017)\citenamefont
  {Slussarenko}, \citenamefont {Weston}, \citenamefont {Li}, \citenamefont
  {Campbell}, \citenamefont {Wiseman},\ and\ \citenamefont {Pryde}}]{Slu17}%
  \BibitemOpen
  \bibfield  {author} {\bibinfo {author} {\bibfnamefont {S.}~\bibnamefont
  {Slussarenko}}, \bibinfo {author} {\bibfnamefont {M.~M.}\ \bibnamefont
  {Weston}}, \bibinfo {author} {\bibfnamefont {J.-G.}\ \bibnamefont {Li}},
  \bibinfo {author} {\bibfnamefont {N.}~\bibnamefont {Campbell}}, \bibinfo
  {author} {\bibfnamefont {H.~M.}\ \bibnamefont {Wiseman}},\ and\ \bibinfo
  {author} {\bibfnamefont {G.~J.}\ \bibnamefont {Pryde}},\ }\href
  {https://doi.org/10.1103/PhysRevLett.118.030502} {\bibfield  {journal}
  {\bibinfo  {journal} {Phys. Rev. Lett.}\ }\textbf {\bibinfo {volume} {118}},\
  \bibinfo {pages} {030502} (\bibinfo {year} {2017})}\BibitemShut {NoStop}%
\bibitem [{\citenamefont {Mart\'{\i}nez~Vargas}\ \emph
  {et~al.}(2021)\citenamefont {Mart\'{\i}nez~Vargas}, \citenamefont {Hirche},
  \citenamefont {Sent\'{\i}s}, \citenamefont {Skotiniotis}, \citenamefont
  {Carrizo}, \citenamefont {Mu\~noz Tapia},\ and\ \citenamefont
  {Calsamiglia}}]{vargas2021quantum}%
  \BibitemOpen
  \bibfield  {author} {\bibinfo {author} {\bibfnamefont {E.}~\bibnamefont
  {Mart\'{\i}nez~Vargas}}, \bibinfo {author} {\bibfnamefont {C.}~\bibnamefont
  {Hirche}}, \bibinfo {author} {\bibfnamefont {G.}~\bibnamefont {Sent\'{\i}s}},
  \bibinfo {author} {\bibfnamefont {M.}~\bibnamefont {Skotiniotis}}, \bibinfo
  {author} {\bibfnamefont {M.}~\bibnamefont {Carrizo}}, \bibinfo {author}
  {\bibfnamefont {R.}~\bibnamefont {Mu\~noz-Tapia}},\ and\ \bibinfo {author}
  {\bibfnamefont {J.}~\bibnamefont {Calsamiglia}},\ }\href
  {https://doi.org/10.1103/PhysRevLett.126.180502} {\bibfield  {journal}
  {\bibinfo  {journal} {Phys. Rev. Lett.}\ }\textbf {\bibinfo {volume} {126}},\
  \bibinfo {pages} {180502} (\bibinfo {year} {2021})}\BibitemShut {NoStop}%
\bibitem [{\citenamefont {Sergeevich}\ \emph {et~al.}(2011)\citenamefont
  {Sergeevich}, \citenamefont {Chandran}, \citenamefont {Combes}, \citenamefont
  {Bartlett},\ and\ \citenamefont {Wiseman}}]{Ser11}%
  \BibitemOpen
  \bibfield  {author} {\bibinfo {author} {\bibfnamefont {A.}~\bibnamefont
  {Sergeevich}}, \bibinfo {author} {\bibfnamefont {A.}~\bibnamefont
  {Chandran}}, \bibinfo {author} {\bibfnamefont {J.}~\bibnamefont {Combes}},
  \bibinfo {author} {\bibfnamefont {S.~D.}\ \bibnamefont {Bartlett}},\ and\
  \bibinfo {author} {\bibfnamefont {H.~M.}\ \bibnamefont {Wiseman}},\ }\href
  {https://doi.org/10.1103/PhysRevA.84.052315} {\bibfield  {journal} {\bibinfo
  {journal} {Phys. Rev. A}\ }\textbf {\bibinfo {volume} {84}},\ \bibinfo
  {pages} {052315} (\bibinfo {year} {2011})}\BibitemShut {NoStop}%
\bibitem [{\citenamefont {Ferrie}\ \emph {et~al.}(2013)\citenamefont {Ferrie},
  \citenamefont {Granade},\ and\ \citenamefont {Cory}}]{Fer13}%
  \BibitemOpen
  \bibfield  {author} {\bibinfo {author} {\bibfnamefont {C.}~\bibnamefont
  {Ferrie}}, \bibinfo {author} {\bibfnamefont {C.~E.}\ \bibnamefont
  {Granade}},\ and\ \bibinfo {author} {\bibfnamefont {D.~G.}\ \bibnamefont
  {Cory}},\ }\href {https://doi.org/10.1007/s11128-012-0407-6} {\bibfield
  {journal} {\bibinfo  {journal} {Quantum Inf Process}\ }\textbf {\bibinfo
  {volume} {12}},\ \bibinfo {pages} {611–623} (\bibinfo {year}
  {2013})}\BibitemShut {NoStop}%
\bibitem [{\citenamefont {Shulman}\ \emph {et~al.}(2014)\citenamefont
  {Shulman}, \citenamefont {Harvey}, \citenamefont {Nichol}, \citenamefont
  {Bartlett}, \citenamefont {Doherty}, \citenamefont {Umansky},\ and\
  \citenamefont {Yacoby}}]{Shu14}%
  \BibitemOpen
  \bibfield  {author} {\bibinfo {author} {\bibfnamefont {M.~D.}\ \bibnamefont
  {Shulman}}, \bibinfo {author} {\bibfnamefont {S.~P.}\ \bibnamefont {Harvey}},
  \bibinfo {author} {\bibfnamefont {J.~M.}\ \bibnamefont {Nichol}}, \bibinfo
  {author} {\bibfnamefont {S.~D.}\ \bibnamefont {Bartlett}}, \bibinfo {author}
  {\bibfnamefont {A.~C.}\ \bibnamefont {Doherty}}, \bibinfo {author}
  {\bibfnamefont {V.}~\bibnamefont {Umansky}},\ and\ \bibinfo {author}
  {\bibfnamefont {A.}~\bibnamefont {Yacoby}},\ }\href
  {https://doi.org/10.1038/ncomms6156} {\bibfield  {journal} {\bibinfo
  {journal} {Nat Commun}\ }\textbf {\bibinfo {volume} {5}},\ \bibinfo {pages}
  {5156} (\bibinfo {year} {2014})}\BibitemShut {NoStop}%
\bibitem [{\citenamefont {Sekatski}\ \emph {et~al.}(2017)\citenamefont
  {Sekatski}, \citenamefont {Skotiniotis},\ and\ \citenamefont
  {D\"ur}}]{Sek17}%
  \BibitemOpen
  \bibfield  {author} {\bibinfo {author} {\bibfnamefont {P.}~\bibnamefont
  {Sekatski}}, \bibinfo {author} {\bibfnamefont {M.}~\bibnamefont
  {Skotiniotis}},\ and\ \bibinfo {author} {\bibfnamefont {W.}~\bibnamefont
  {D\"ur}},\ }\href {https://doi.org/10.1103/PhysRevLett.118.170801} {\bibfield
   {journal} {\bibinfo  {journal} {Phys. Rev. Lett.}\ }\textbf {\bibinfo
  {volume} {118}},\ \bibinfo {pages} {170801} (\bibinfo {year}
  {2017})}\BibitemShut {NoStop}%
\bibitem [{\citenamefont {Tonekaboni}\ \emph {et~al.}(2022)\citenamefont
  {Tonekaboni}, \citenamefont {Chantasri}, \citenamefont {Song},\ and\
  \citenamefont {Wiseman}}]{PRA}%
  \BibitemOpen
  \bibfield  {author} {\bibinfo {author} {\bibfnamefont {B.}~\bibnamefont
  {Tonekaboni}}, \bibinfo {author} {\bibfnamefont {A.}~\bibnamefont
  {Chantasri}}, \bibinfo {author} {\bibfnamefont {H.}~\bibnamefont {Song}}, \bibinfo {author} {\bibfnamefont {Y.}~\bibnamefont
  {Liu}},\
  and\ \bibinfo {author} {\bibfnamefont {H.~M.}\ \bibnamefont {Wiseman}},\
  }\href@noop {} {\bibfield  {journal} {\bibinfo  {journal} {arXiv:2205.12566
  [quant-ph]}\ } (\bibinfo {year} {2022})}\BibitemShut {NoStop}%
\bibitem [{\citenamefont {Zorin}\ \emph {et~al.}(1996)\citenamefont {Zorin},
  \citenamefont {Ahlers}, \citenamefont {Niemeyer}, \citenamefont {Weimann},
  \citenamefont {Wolf}, \citenamefont {Krupenin},\ and\ \citenamefont
  {Lotkhov}}]{ZorAhl1996}%
  \BibitemOpen
  \bibfield  {author} {\bibinfo {author} {\bibfnamefont {A.~B.}\ \bibnamefont
  {Zorin}}, \bibinfo {author} {\bibfnamefont {F.-J.}\ \bibnamefont {Ahlers}},
  \bibinfo {author} {\bibfnamefont {J.}~\bibnamefont {Niemeyer}}, \bibinfo
  {author} {\bibfnamefont {T.}~\bibnamefont {Weimann}}, \bibinfo {author}
  {\bibfnamefont {H.}~\bibnamefont {Wolf}}, \bibinfo {author} {\bibfnamefont
  {V.~A.}\ \bibnamefont {Krupenin}},\ and\ \bibinfo {author} {\bibfnamefont
  {S.~V.}\ \bibnamefont {Lotkhov}},\ }\href
  {https://doi.org/10.1103/PhysRevB.53.13682} {\bibfield  {journal} {\bibinfo
  {journal} {Phys. Rev. B}\ }\textbf {\bibinfo {volume} {53}},\ \bibinfo
  {pages} {13682} (\bibinfo {year} {1996})}\BibitemShut {NoStop}%
\bibitem [{\citenamefont {Paladino}\ \emph {et~al.}(2002)\citenamefont
  {Paladino}, \citenamefont {Faoro}, \citenamefont {Falci},\ and\ \citenamefont
  {Fazio}}]{PalFao2002}%
  \BibitemOpen
  \bibfield  {author} {\bibinfo {author} {\bibfnamefont {E.}~\bibnamefont
  {Paladino}}, \bibinfo {author} {\bibfnamefont {L.}~\bibnamefont {Faoro}},
  \bibinfo {author} {\bibfnamefont {G.}~\bibnamefont {Falci}},\ and\ \bibinfo
  {author} {\bibfnamefont {R.}~\bibnamefont {Fazio}},\ }\href
  {https://doi.org/10.1103/PhysRevLett.88.228304} {\bibfield  {journal}
  {\bibinfo  {journal} {Phys. Rev. Lett.}\ }\textbf {\bibinfo {volume} {88}},\
  \bibinfo {pages} {228304} (\bibinfo {year} {2002})}\BibitemShut {NoStop}%
\bibitem [{\citenamefont {Gardiner}(1985)}]{Gar85}%
  \BibitemOpen
  \bibfield  {author} {\bibinfo {author} {\bibfnamefont {C.~W.}\ \bibnamefont
  {Gardiner}},\ }\href@noop {} {\emph {\bibinfo {title} {Handbook of Stochastic
  Methods}}}\ (\bibinfo  {publisher} {Spring\-er},\ \bibinfo {address}
  {Berlin},\ \bibinfo {year} {1985})\BibitemShut {NoStop}%
\end{thebibliography}

%

\end{document}